# Magnetization and Hall effect studies on the pyrochlore iridate $Nd_2Ir_2O_7$

S. M. Disseler[1], S. R. Giblin[2], Chetan Dhital[1], K. C. Lukas[1], Stephen D. Wilson[1], and M. J. Graf[1]

[1] Department of Physics, Boston College, Chestnut Hill, MA 02467 USA
[2] School of Physics and Astronomy, Cardiff University, Cardiff CF24 3AA, UK

Abstract

We present magnetization and Hall effect measurements on the pyrochlore iridate $Nd_2Ir_2O_7$. Previous muon spin rotation measurements have shown that the system undergoes an unusual transition at $T_M \approx 110$ K into a magnetic phase lacking long-range order, followed by a transition at $T_{LRO} \approx 6$ K into a state with long-range magnetic order. We observe a small remnant magnetization when cycling through zero magnetic field at temperatures below $T_M$. Below $T_{LRO}$, this remnant magnetization increases sharply, and new hysteresis effects appear at a higher field $B_c = 2.5$ T, while the Hall resistance develops a non-monotonic and hysteretic magnetic field dependence, with a maximum at $B_c$ and signatures of an anomalous Hall effect. The dependence on field sweep direction demonstrates a non-trivial transition into a magnetically ordered state with properties paralleling those of known spin-ice systems and suggests a spin reorientation transition across the metal insulator transition in the A-227 series.

The pyrochlore iridate $A_2Ir_2O_7$(A-227, A=Y, RE) is an especially appealing system as it allows for the study of the interplay of SOC, electronic correlations, and band topology. This family has been predicted to host exotic phases including axion insulators, strong topological insulators, and a Weyl semi-metal[1-3]. However, aside from the more exotic topological phases predicted in the electronic phase diagram, magnetic frustration also plays a role in the groundstate of this system, where both the $Ir^{4+}$ and $A^{3+}$ ions may be magnetic with each occupying interpenetrating networks of corner-sharing tetrahedra. This potential for magnetic frustration and a corresponding quantum spin liquid ground state has driven a parallel search for signatures of spin order reflecting the well-studied manifold of spin states on geometrically frustrated tetrahedra intrinsic to spin-ice compounds. Realizing such a state in the presence of strong spin-orbit coupling presents the potential for a host of novel electronic phenomena such as the manifestation of anomalous quantum Hall effects induced via the presence of finite chirality in a spin-liquid state.

Such a state has recently been proposed within the metallic pyrochlore iridate $Pr_2Ir_2O_7$ where a chiral spin liquid phase is proposed to drive an anomalous Hall response in the absence of conventional, long-range, spin order [4]. How such a response evolves as the bandwidth of this material is tuned and screening/electron hopping effects are reduced however remains unexplored. Specifically, one key question remains: whether the proposed chiral spin liquid phase of Pr-227 evolves continuously into the Mott insulating ground state of the smaller bandwidth A-227 systems (e.g. $Y_2Ir_2O_7$, $Eu_2Ir_2O_7$, $Yb_2Ir_2O_7$) [5,6], or whether magnetic interactions are renormalized dramatically near the phase boundary to the insulating regime.

In order to perturb the proposed chiral spin liquid ground state of Pr-227 via enhanced correlations, one of the most promising avenues is therefore exploring the weakly metallic variants of the $Nd_2Ir_2O_7$ compound, where naively this system is in very close proximity to the spin-liquid ground state of Pr-227. Previously it has been shown that the low energy magnetic and electronic states are related to the A-site radii, which set the scale for electron-electron correlations via tuning their resulting bandwidth [7, 8, 9]. Specifically, A-227 iridates composed of the smallest A-site radii (A= Y, Yb-Eu) are insulating for all temperatures, whereas the compound with the largest A-site radius, Pr-227, remains metallic down to 30 mK. This leaves Nd-227 as a transitional compound within the A-227 series where the resulting ground state (*e.g.* insulating [10] or weakly metallic [11]) is highly sensitive to the sample stoichiometry.

Our recent measurements on high quality polycrystalline samples of Nd-227 [11] revealed weakly metallic behavior which crossed over below 6 K to a logarithmically increasing resistivity with decreasing temperature, as expected for Kondo systems, an effect also observed in Pr-227 [12]. The observation of a negative longitudinal magnetoresistance that varies quadratically with magnetization further strengthens the argument for a Kondo-like mechanism [11]. A combination of magnetic susceptibility and zero-field muon spin relaxation studies of this same weakly metallic Nd-227 system revealed short-range/fluctuating order below 110 K with long range order unambiguously observed only below ~ 6 K. This demonstrates that the magnetic interactions are appreciably strengthened by the shift in the A-site ion from $Pr^{3+}$ to $Nd^{3+}$, where for Pr-227 the Pr sublattice enters the proposed chiral spin-liquid state [4] below the Curie-Weiss temperature $\theta_{CW}$ ~1.7 K; based on the anisotropic response of the magnetization the local spin structure is inferred to be 2-in/2-out (2I2O). The resulting complex, weakly metallic, ground state in Nd-227, which preserves a sizeable moment on the RE-site and the accompanying f-d exchange interactions, is therefore an ideal setting for probing how increased correlation effects influence the reported chiral spin liquid phase of Pr-227. Specifically, exploring how the all-in/all-out (AIAO) type spin order proposed for insulating Nd-227 variants [13] compares to our weakly metallic Nd-227 samples may hold important clues for understanding the interactions responsible for the exotic electronic behavior observed in the far metallic Pr-227 region of the A-227 phase diagram.

In this communication we present detailed measurements of the field dependent hysteresis of bulk magnetization and the Hall effect in poorly metallic Nd-227 whose local field behavior is already known via previously reported µSR measurements. We find a non-zero remnant magnetization on the order of 0.01 µ$_B$/Ir below $T_M$ = 110 K, a temperature coinciding with the bifurcation in magnetic susceptibility observed in our prior work. Below $T$ ~ 10 K, a second hysteresis centered about a critical field $B_c$ ~ 2.5 T is also observed, coinciding with a sharp increase in the zero-field remnant magnetization. In addition to the large negative magnetoresistance below 10 K, we find a striking non-monotonic Hall effect that shows considerable hysteresis peaked at $B_c$. Our results provide evidence that the spin behavior and electronic response of Nd-227 manifests as a precursor phase intermediate between the long-range ordered Eu-227 and the disordered chiral spin liquid of Pr-227. Furthermore, our data suggest that the "all-in/all-out" spin configuration previously reported in more insulating Nd-227 variants reorients into a "two-in/two-out" spin ice configuration across the metal-insulator transition (MIT) within the A-227 series.

Polycrystalline samples of $Nd_2Ir_2O_7$ were synthesized by reacting stoichiometric amounts of $Nd_2O_3$ (99.99%) and $IrO_2$ (99.9%) as previously described [11]. Magnetization measurements were performed in a Quantum Design MPMS SQUID magnetometer. Data for isothermal hysteresis loops were taken by cooling the sample in nominal zero field, and sweeping the field over the cycle 0 → 5 → -5 → 0.1 or 5 T. Individual data points were taken while sweeping magnetic field at a constant rate of 70 mT/min. The sample was warmed to $T$ = 130 K > $T_M$ after each isothermal run. Magnetotransport was measured at and above 2 K in a Quantum Design PPMS system, and below 2 K in a $^3$He cryostat with a 9 T magnet. Several pressed samples prepared from independent batches were studied; all were polished into bar geometries with typical sample with thicknesses in the range 0.5 – 1 mm. For Hall effect measurements, two contacts were placed at sample edges perpendicular to both the current flow and field directions, while for longitudinal magnetoresistance measurements the Hall voltage was shorted with the voltage leads extending across the sample, and the magnetic field was applied parallel to the current direction.

In Fig. 1 we show the magnetization at $T$ = 2 K over the field cycle 0 → 5 → -5 → 0.1 T. Data are presented as $\mu_B$ per Ir, although our earlier work [11] showed that paramagnetic $Nd^{3+}$ makes a large contribution to the total magnetization. A small hysteresis loop centered on zero field (lower right inset) is seen with a zero-field moment of ±0.015$\mu_B$/Ir, roughly three orders of magnitude larger than that observed for Y-227 in the long-range ordered state near 100 K as reported in Ref. 14, and two orders of magnitude larger than measured by us on our own Y-227 samples [15]. Also evident is a hysteresis loop centered on a field value $-B_c \approx -2.5$ T. The magnitude of the hysteresis is 2.5 times larger than that at zero field. Measurements over the field cycle $0 \rightarrow 5 \rightarrow -5 \rightarrow 5$ T at $T = 3, 4$, and 5 K show that the hysteresis occurs at both positive and negative values of $B_c$. Plotting the derivative $dM/dB$ for the up and down field sweeps (upper left inset) shows that a peak occurs at a field of magnitude $B_c$ and pointing opposite to the previous magnetizing field direction, that is, at $+B_c$ for the field sweep -6 to 6 T, and at $-B_c$ for 6 to -6 T. The virgin curve $0 \rightarrow 6$T also shows a peak in dM/dB at $B_c$ (not shown), but with a size about one-half those depicted in the inset. This behavior is not typical for metamagnetic transitions (*e.g.* spin-flip or spin-flop), which typically occur at $\pm B_c$ for both sweep directions, with potentially some small hysteresis about that value.

In Fig. 2 we have plotted the temperature dependence of the difference between the down and up sweep magnetizations, $\Delta M = M_{down}(B) - M_{up}(B)$ for zero field ($\Delta M_{ZF}$ in Fig. 2a) and at $B_c$ ($\Delta M_{HF}$ in Fig. 2b); the insets show representative curves taken at several temperatures. In Fig. 2a we see two abrupt changes in $\Delta M_{ZF}$, near 110 K and 10K. These correspond closely to the two transitions observed in our earlier µSR, transport, and dc susceptibility results [12], $T_M \approx 110$ K and $T_{LRO} \approx 6$ K, corresponding to the onset of short-and long-range magnetic order, respectively. We also note that two transitions were observed in Y-227, with $T_M \approx 190$ K and $T_{LRO} \approx 150$ K [5], and the results for the temperature dependence of the remnant moment reported in Ref. [15] on that system are similar to our results shown in Fig. 2a and consistent with those transition temperatures.

From Fig. 2b we see that as the temperature decreases below 10 K, $\Delta M_{HF}$ increases rapidly from zero, and so we assume the onset of hysteresis at $B_c$ is associated with the onset of long-range magnetic order at $T_{LRO}$. A high field hysteresis was observed in Pr-227 below the spin freezing temperature for fields applied along the [111] direction only [4], however as full hysteresis loops have not been published on that system, it is unknown if a similar behavior with regard to field sweep direction exists in Pr-227 as well. Regardless, it was estimated, based on the size of the localized $Pr^{3+}$ moments, that the hysteresis corresponded to a partial stabilization of 3-in/1-out (3I3O) within the 2-in/2-out (2I2O) matrix below the spin freezing temperature, consistent with Monte Carlo simulations incorporating RKKY interactions on the pyrochlore lattice [16]. Measurements of the magnetization of the insulating spin ice compound $Ho_2Ti_2O_7$ at low temperatures have also shown transitions from 2I2O to 3I1O at high fields applied along the [111] direction, which are hysteretic in nature [17]. We have also made a preliminary measurement of the magnetocaloric effect on Nd-227 at $T = 0.5$ K [15], and find a large heat release at $B_c$ which has the same field history dependence as shown in the upper inset of Fig. 1. We conclude that the magnetization hysteresis at $B_c$ is the signature of entry into a lower entropy magnetically ordered state, e.g., a 2I2O to 3I1O or AIAO transition.

We now describe our Hall effect results. Due to imperfect alignment and probable current flow irregularities in the sintered samples, our signal had a large contribution from the resistance (Fig. 3a). The Hall voltage was extracted by taking the antisymmetric part of the measured voltage and then examining the field history dependence. This was done for field sweeping from -6 to 6 T and also from 6 to − 6 T. At 10 K the magnetic field response is dominated by the Hall contribution because the resistance is essentially field independent, but at lower temperatures the magnetoresistance is large [11], and so the Hall (antisymmetric) component is only 4% of the total change. Because the Hall signal is a relatively small fraction of the total signal, we independently measured the longitudinal magnetoresistance [15]; no asymmetry is observed with respect to field, and any hysteresis observed between sweep directions near $B_c$ was near the limits of the experimental error. We note that the offset at zero field between the virgin zero field cooled and zero field after ramping to 6 T was observed in both Hall and longitudinal geometries.

The field dependence of the Hall resistivity is shown in Fig. 3b with each isotherm offset for clarity. While at temperatures higher than 10 K the signal is linear and nearly independent of field sweep direction, at lower temperatures we find a non-monotonic variation of the Hall resistivity with field, with strong field history dependence. In Fig 3c and 3d we show the reversible and irreversible components of the field dependence of $\rho_{xy}$; these are found by taking half of the sum (reversible) or difference (irreversible), of the forward and backward sweep directions shown in Fig. 3b for each temperature. The magnetic field dependence of the reversible component is characteristic of an anomalous Hall effect, which is expected in a system with magnetic ordering and strong spin-orbit coupling, as discussed below. In this case, we can find the approximate carrier concentration by fitting the reversible part of the data at high fields ($B > 4$ T) where $M$ has nearly saturated, allowing us extract a lower limit on $R_0$. Using the single band approximation, $R_0 = -1/ne$, and we find an upper limit of the carrier concentration of $2.5 \times 10^{20}$ cm$^{-3}$, compared to $4 \times 10^{21}$ cm$^{-3}$ measured in small single crystals of Pr-227 [18]. This confirms our expectation that Nd-227 exhibits properties intermediate between those of metallic Pr-227 and the insulating members of the A-227 family. Following the arguments of Ref. 18, and assuming a spherical Fermi surface with $k_F = (3\pi^2 n)^{1/3}$ we find the resultant RKKY interaction between neighboring Nd$^{3+}$ ions to be ferromagnetic, as in the case of Pr-227.

These results are remarkably similar to that obtained in artificial spin-ice structures [19] for which the field dependence of the reversible component of the Hall resistance is that typically associated with an AHE in ferromagnetic metals [20], and the irreversible component develops a peak about a critical field $B_c$ with decreasing temperature. Magnetization extracted from magnetotransport in Ref 19 shows a sharp change in dM/dH about the critical field, which also depends on field sweep direction in a manner similar to what we have observed in Nd-227. These features were attributed to a spatially varying chirality around each hexagonal loop in the Kagome lattice studied in that work, and therefore may be directly related to phenomena observed in the conducting pyrochlore compounds which have a spin-ice configuration.

The similarities between the field dependence of the magnetization and Hall effect shown here for Nd-227 and those of other spin-ice compounds strongly imply that the magnetic ground state of Nd-227 in the weak metallic phase is of the 2I2O spin-ice type rather than AIAO as previously considered for more insulating Nd-227 variants. Recent theoretical work incorporating f-d exchange supports a 2I2O ground state for a wide range of coupling parameters, particularly if the average (CurieWeiss) interaction between neighboring Nd$^{3+}$ is ferromagnetic [21]. The spin-ice state in the iridates has theoretically been shown to exhibit a transition into a number of more ordered phases at high magnetic fields including the Kagome-ice state [22], 3I1O, and even AIAO [23], depending on the conduction electron density and strength of distant-neighbor interactions. It has also recently been predicted that spin-ice structures may give rise to resistivity minimum as in traditional Kondo systems [24], and in fact it was shown [25] to agree with previous data for single crystal samples made metallic though applied external pressure [26]. To our knowledge it has not been demonstrated either theoretically or experimentally that any of these phenomena will occur if the system is in an AIAO ground state.

Despite these similarities, the presence of LRO in zero field as observed in μSR measurements is inconsistent with that of traditional spin-ice behavior and the chiral spin liquid state proposed within Pr-227. One possible explanation for this behavior is that the onset of hysteresis at zero field below $T_M$ and near $B_c$ below $T_{LRO}$ correspond to distinct order parameters, such as order of the individual sublattices mediated by *d-d*, *f-f* or *f-d* interactions. We note that these two temperatures are similar to that found in the Yb-227, in which the $Ir^{4+}$ moments order at $T_{LRO}$ = 125 K, while the localized paramagnetic $Yb^{3+}$ moments become polarized due to this local field and interact with an effective Curie-Weiss temperature of $\theta_{CW}$ ~ –7 K [6], with probable ordering in the Yb sublattice below 3 K [15]. Therefore with the addition of the RKKY interaction in Nd-227, we expect there should be significantly more impact beyond simple polarization on the underlying ground state of the Ir- sublattice and vice-versa. It is also possible that at low temperatures and zero magnetic field the ordering is incomplete, producing an electronically phase-separated heterogeneous state composed of domains long-range magnetic order imbedded within a matrix of short-range ordered material, as proposed for the Pr-227 [4]. Indeed, our μSR measurements [11] show that the depolarization in zero field does not follow the form typically associated with simple LRO as seen in the insulating A-227 compounds [5,6], requiring an additional component which could be associated with the domains of short-range order.

In summary, we have studied high quality polycrystalline Nd-227 via magnetization and Hall effect. Our combined results reveal a complex precursor phase in the A-227 series at the boundary between the proposed metallic spin liquid and the long-range AF ordered insulating phase where a signatures of time reversal symmetry breaking emerge prior to the onset of static spin order. This intermediate state manifests electronic behavior reflecting the presence of two field and temperatures scales, suggesting two order parameters in this system.The striking similarities of our results to those for other spin-ice systems suggests that Nd-227 in the weakly metallic regime has a two-in/two out rather than all-in/all-out magnetic structure potentially implying a spin reorientation transition across the MIT in the A-227 series.

We would like to acknowledge the technical assistance of Tom Hogan. This work was supported in part by National Science Foundation Materials World Network grant DMR-0710525 (M.J.G.) and by NSF CAREER award DMR-1056625 (S.D.W.).Magnetization experiments were performed at the ISIS Materials Characterization Laboratory at the RutherfordAppleton Laboratories (UK).

**Figure Captions**

Figure 1. Magnetization hysteresis loop for $Nd_2Ir_2O_7$, with a field cycle of 0 → 5 → -5 → 0.1T. Lower right inset: close-up of the hysteresis in the vicinity of zero field. Upper left inset: Derivative *dM/dH* (in units of $10^{-2} \mu_B/T$) for 5 → -5T (lower curve) and -5 to 5 T (upper curve), taken at $T = 3$ K.

Figure 2. (a) Temperature variation of the maximum in $\Delta M = M_{down} - M_{up}$ at zero field. The inset shows detailed curves for $\Delta M$ in the vicinity of zero field at three temperatures. (b) Temperature variation of maximum in $\Delta M$ near -2 T. The inset shows typical curves for $\Delta M$ at three temperatures. Different symbols (solid circles, solid triangles) represent data taken on different experimental runs.

Figure 3. (a) Resistance versus magnetic field for a sample in the transverse geometry over the cycle 6 → -6 → +6 T at several temperatures. Inset: Resistivity (in units of $\mu\Omega$ cm) versus temperature in 0 and 9 T fields, as reported in Ref. X. (b) Hall resistivity extracted from the antisymmetric part of the curves in Fig. 3a from 6 → - 6 T (open symbols, left-pointing arrows) and -6 → 6 T (closed symbols, right-pointing arrows). The curves at 4 and 10 K are offset by 0.11 and 0.22 $\mu\Omega$ cm, respectively, for clarity.

Figure 1

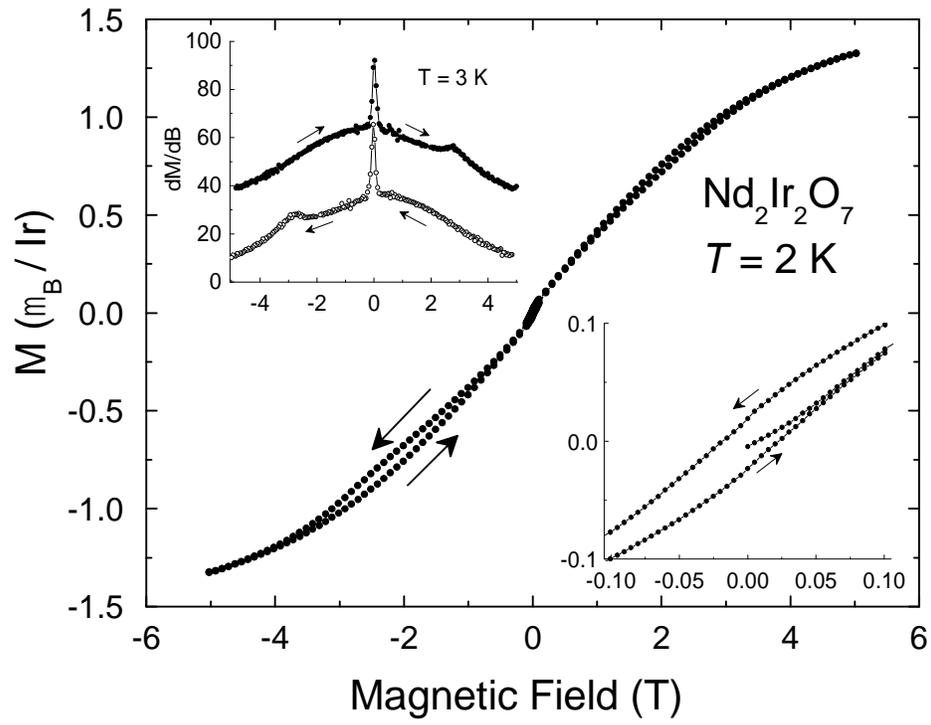

Figure 2

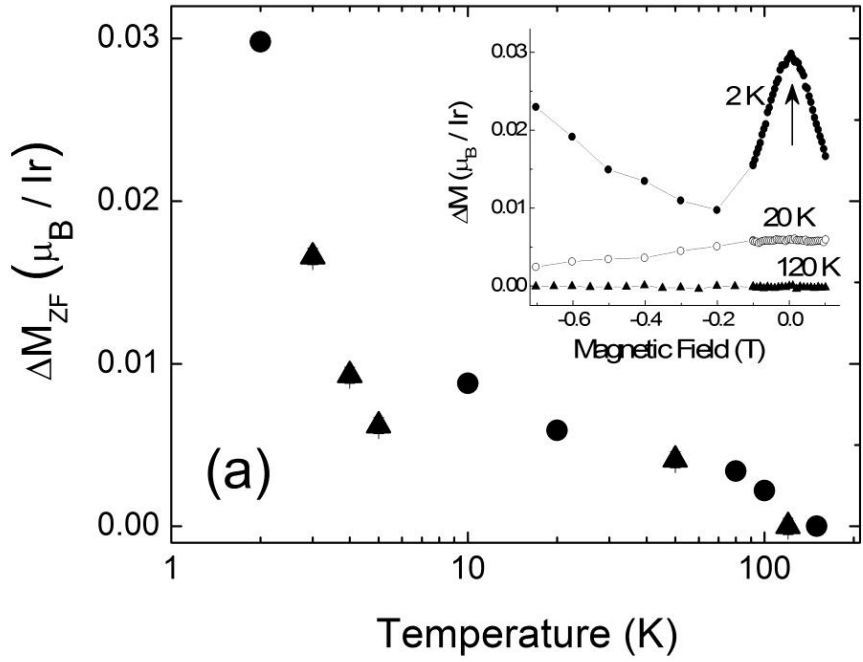

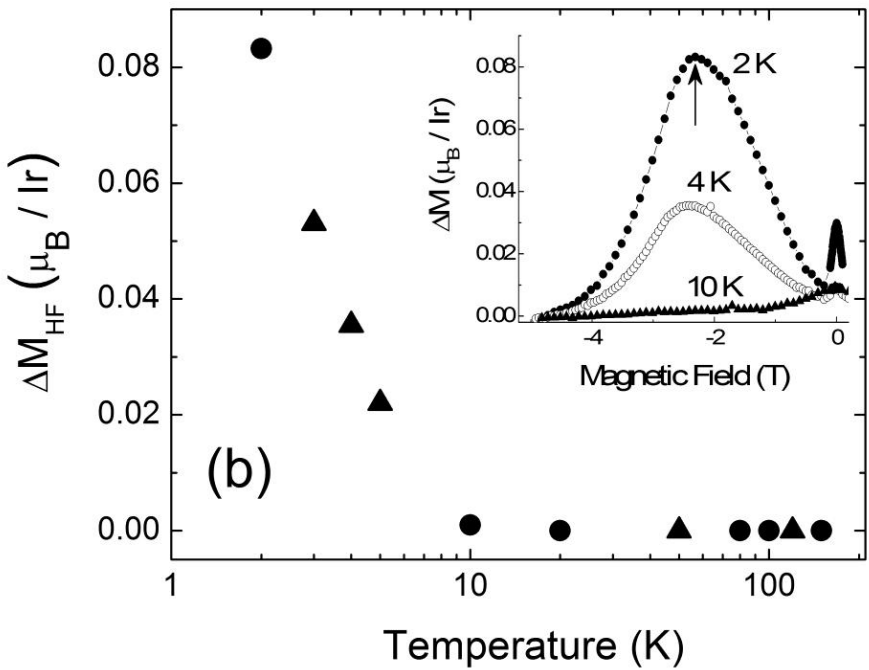

Figure 3

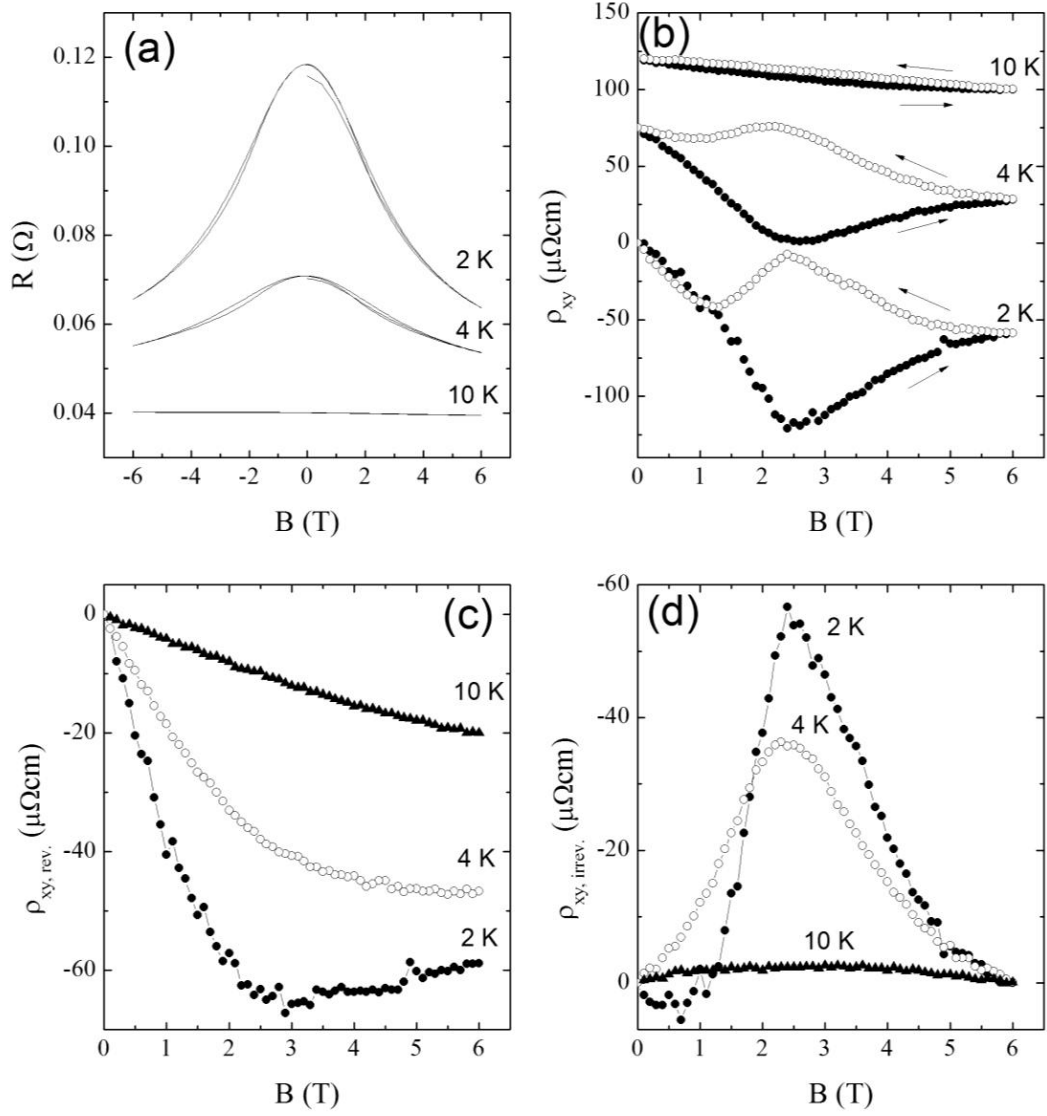

# Supplemental Information

Magnetization and Hall effect studies on the pyrochlore iridate $Nd_2Ir_2O_7$

S. M. Disseler *et al*.

**Magnetoresistance of $Nd_2Ir_2O_7$**

In order to discern effects arising from the magnetoresistive and Hall components, we have measured the longitudinal magnetoresistance of Nd-227. The current was applied parallel to the magnetic field, and silver epoxy was used to attach leads and to attempt to short any transverse voltages occurring to do misalignment relative to the field. The normalized Hall and longitudinal resistances are computed as (R(H)-R(0))/R(0) and are shown in Fig. S1 for data taken at 500 mK. We note that both components show a sizeable difference between the virgin zero field and after one half of the completed loop. The mechanism for this feature is currently unknown and is the subject of future investigations. There is no asymmetry with regards to the sign of the field, however we note that a very small hysteresis, less than one-fifth of that seen in the Hall component, is observed here.

**Magnetization of $Y_2Ir_2O_7$**

Previous transport, dc-susceptibility, neutron scattering, and μSR results for Y-227 showed the dc magnetization anomaly at $T_M$ = 190 K, followed by the onset of long-range magnetic order at $T_{LRO}$ = 150 K [1]. A limited number of magnetization loops were measured; for each isothermal measurement, samples were warmed to $T$ > 200 K and then cooled in zero field to the corresponding measurement temperature. As shown in Fig. S2, above $T_M$ = 190 K simple paramagnetic behavior is observed, while below $T_M$ a narrow hysteresis with remnant moment ~ 1 x $10^{-3} \mu_B/Ir^{4+}$ is found. At 2 K (below $T_{LRO}$) the remnant moment has approximately doubled, while the coercive field has increased by more than an order of magnitude to nearly 10 kOe. However, no additional hysteretic effects are observed at higher fields as observed in Nd-227.

**Magnetocaloric Effect in Nd$_2$Ir$_2$O$_7$**

A preliminary measurement of the magnetocaloric effect (MCE) was performed on Nd-227 at 500 mK in a sample-in-vacuum $^3$He cryostat. The sample was in a high vacuum environment, and isolated from the $^3$He bath by 5mm of Stycast 1266 epoxy; a bare-chip Cernox 1030 thin film thermometer was attached to the sample with conductive silver paint, and 2-mil brass wires approximately 5 mm in length acted as electrical connections and as weak thermal links to the bath. A second factory-calibrated Cernox 1030 thermometer was used to measure the temperature of the cryostat during the field sweep and to calibrate the bare chip thermometer attached to the sample. The field was then swept from +6 T → -6 T and then from -6 T → +6 T at a constant rate of 0.045 mT/min.

Shown in Fig. S3, the sample temperature increases greatly near ± $B_c$ depending on sweep direction, indicating a release of heat by the sample. The field history dependence of the MCE follows closely with that observed in magnetization and Hall effect as presented in the main text of the article, from which we conclude that the sample enters a lower entropy state at high fields, causing the release of heat near $B_c$. Any intrinsic effects near zero magnetic field are obscured by magnetocaloric effects related to magnetic impurities in the copper platform, because these effects are systematically observed in the bath thermometer as well. We also note that a drift in the base temperature of the cryostat during the positive sweep direction caused the apparent offset between of the up and down sweep directions, however this does not qualitatively affect the results presented here.

**Temperature dependence of the local field in Yb$_2$Ir$_2$O$_7$**

Previously, zero-field μSR measurements were performed to 2 K on sample of Yb-227 (shown as open circles in the Fig. S4), and these revealed that (1) $T_M = T_{LRO} =$ 130 K, and (2) the ordered Ir$^{4+}$ sublattice polarizes the local Yb$^{3+}$ moments which interact with an effective Curie-Weiss temperature θ$_{CW}$ ~ - 7 K [1]. We recently conducted additional zero-field measurements at Paul Scherrer Institute utilizing the Low Temperature Facility to extend the temperature range to 20 mK. The spontaneous muon precession frequency, shown in Fig. S4, was found by taking the Fourier transform of the time domain asymmetry data, and fitting a Lorentzian peak distribution to the resulting spectra. We find a small peak in the temperature-dependent precession frequency near 1.6 K, indicating either full polarization of the $Yb^{3+}$ moments or the onset of magnetic order within the Yb sublattice, which slightly decreases the net magnetic field at the muon stopping site, reaching a constant value in the low temperature limit as shown in the inset of Fig S4. We note that several data points were taken at the lowest temperature, and all give similar frequency values with the experimental resolution, indicating the sample was in thermal equilibrium near the base temperature.

We would like to acknowledge the technical assistance of Alex Amato and Chris Baines for Yb-227 data taken at PSI.

Supplemental Figure S1

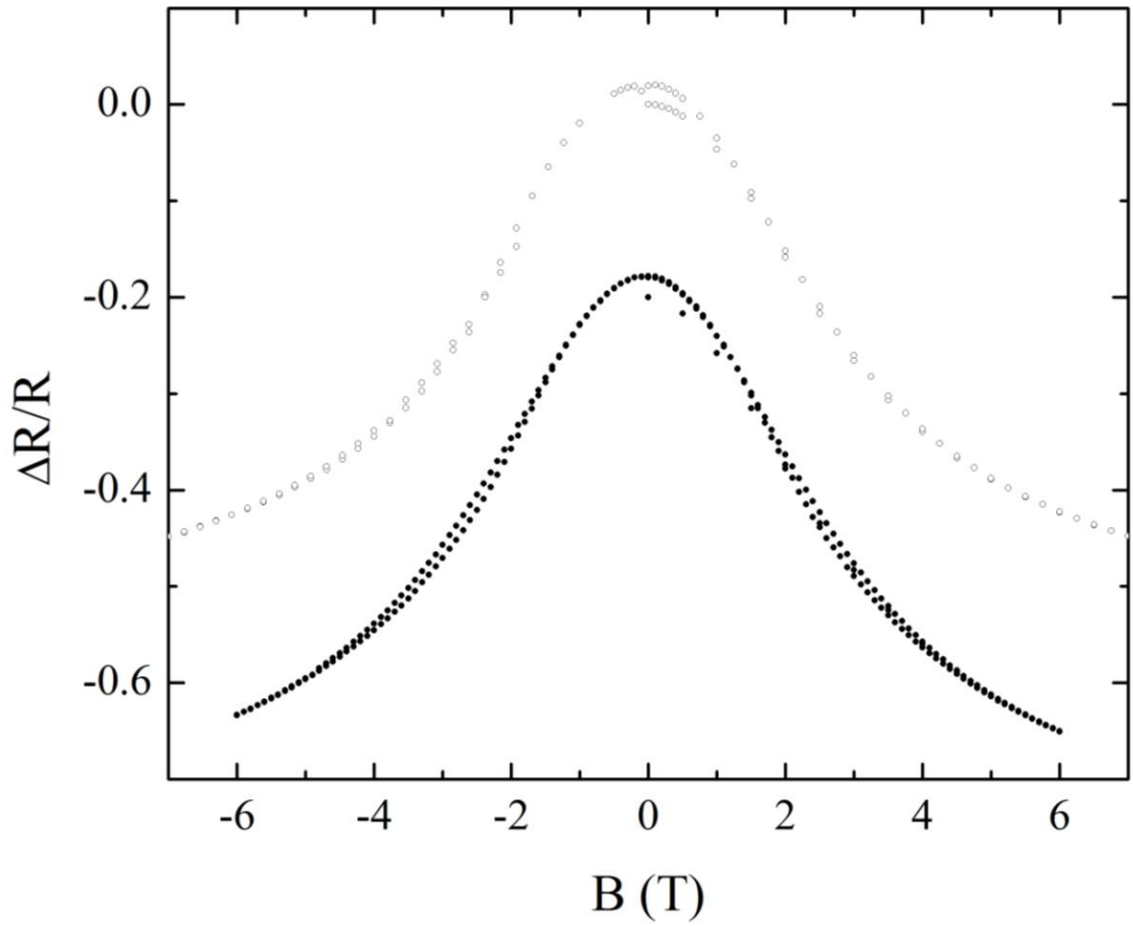

Supplemental Figure S1: Normalized magnetoresistance for Hall (closed circles) and longitudinal geometries ( open circles). The Hall data has been offset for clarity.

Supplemental Figure S2

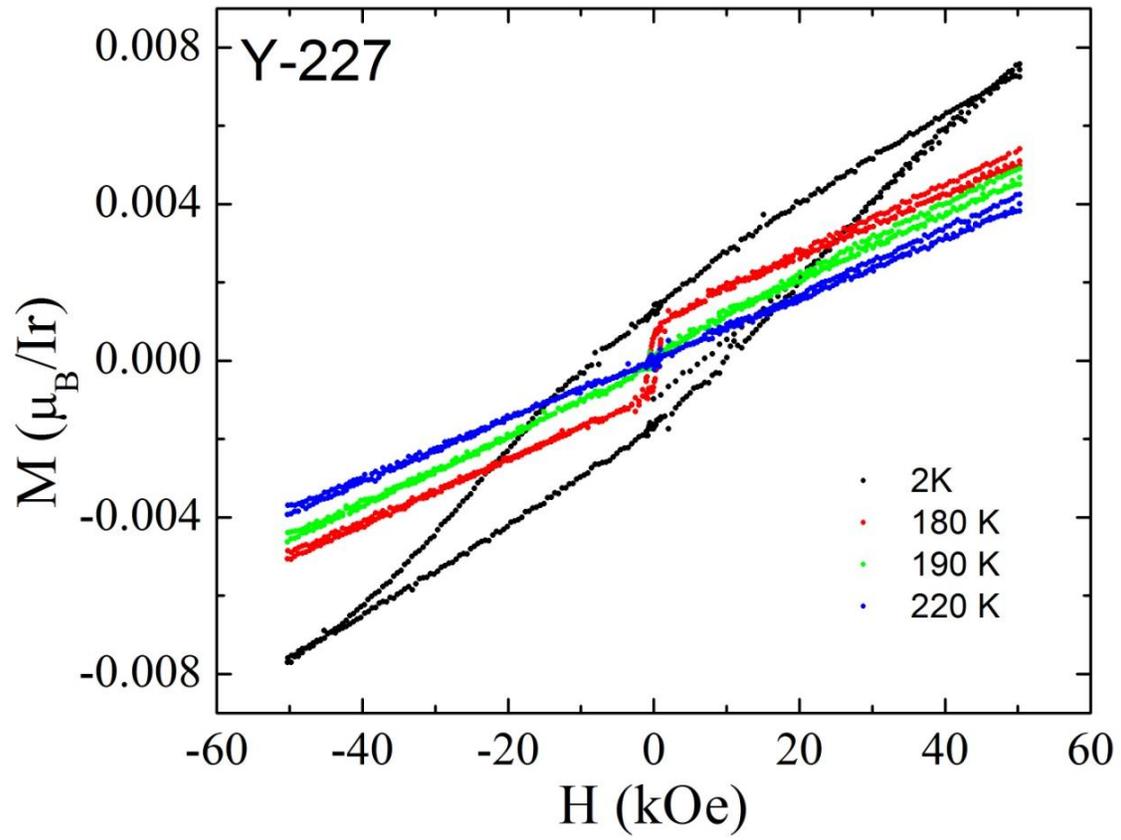

Supplemental Figure S2. Magnetization versus field isotherms taken on Y-227.

Supplemental Figure S3

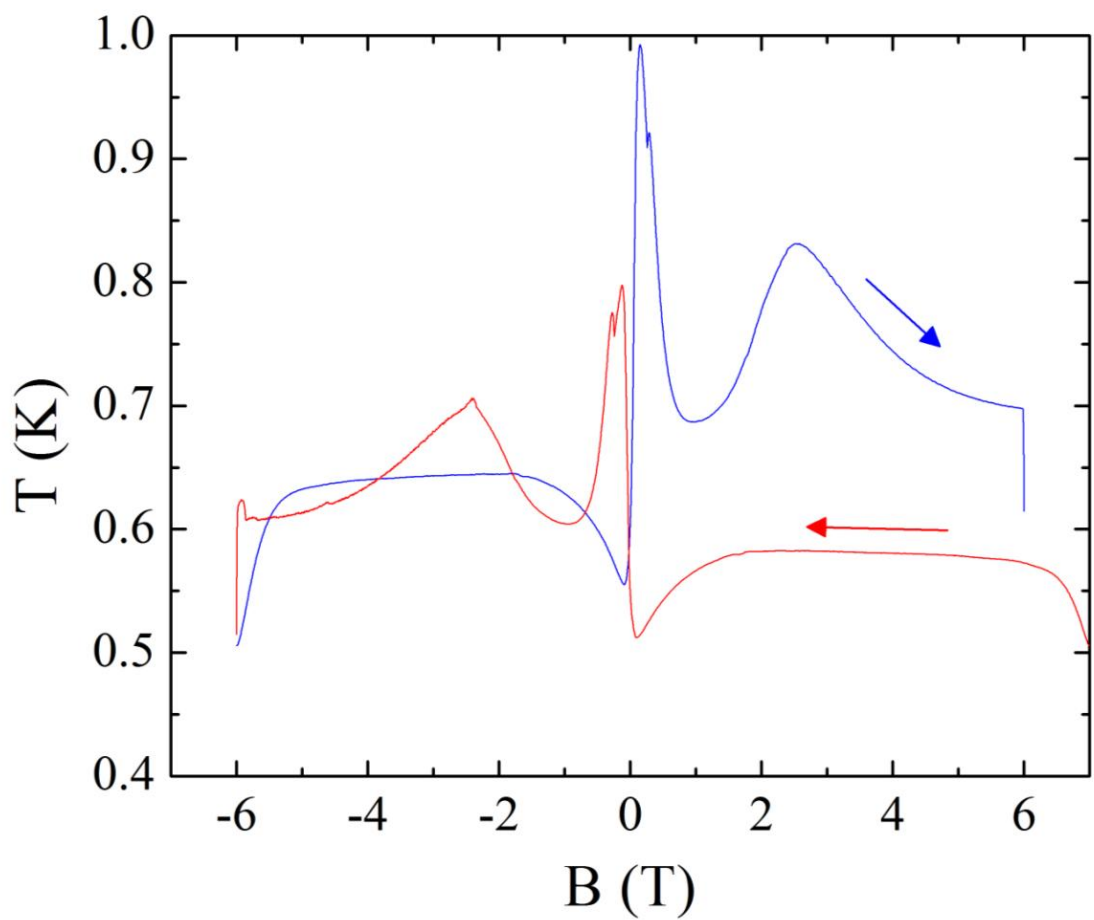

Supplemental Figure S3. Temperature of the sample showing the MCE for field swept at a constant rate of 0.045 mT/min. The red curve and blue curves are for the negative (+6 T to -6 T), and positive (-6T to +6 T) sweep directions respectively.

Supplemental Figure S4

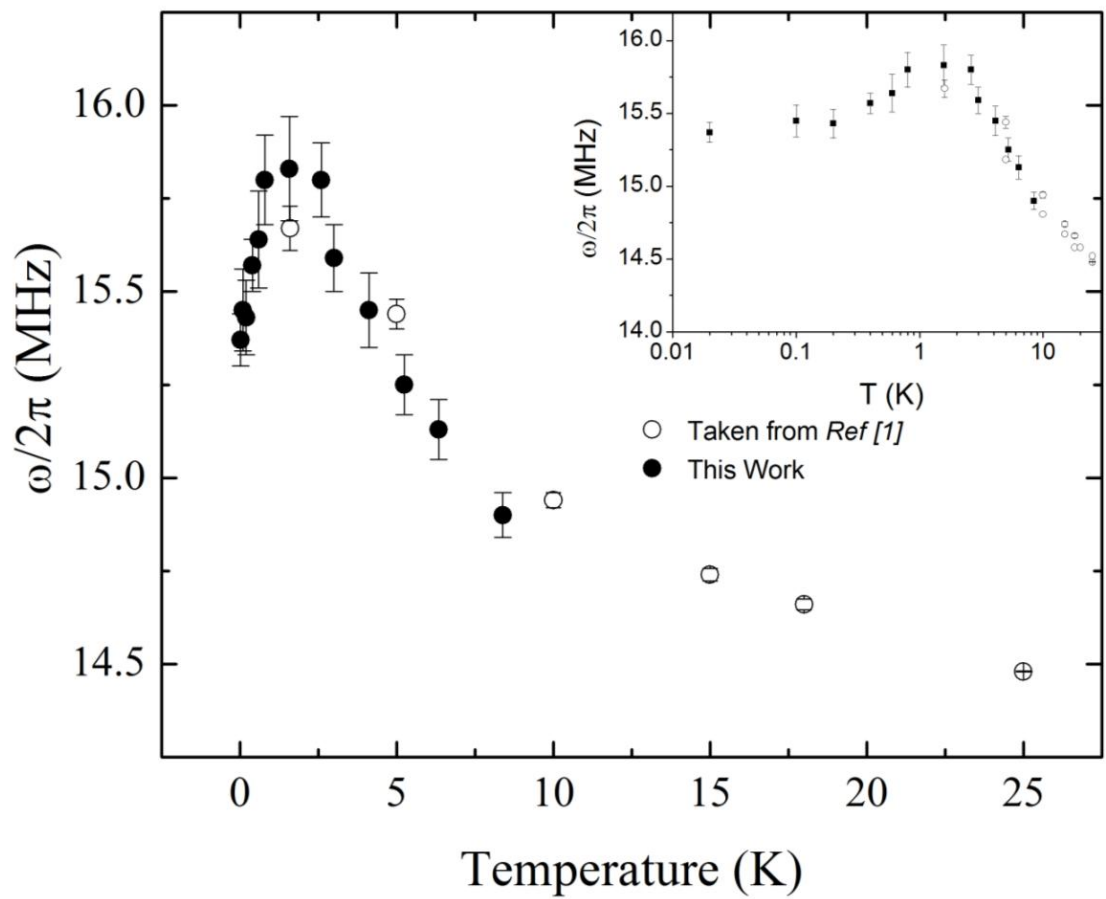

Supplemental Figure S4. Temperature dependence of the spontaneous oscillation frequency. Data taken from Ref 6 (see main text) is shown as open circles, while new results are shown as closed circles. Inset: data shown on logarithmic temperature scale.